\documentclass[graybox]{svmult}

\usepackage{type1cm}        % activate if the above 3 fonts are
\usepackage{makeidx}         % allows index generation
\usepackage{graphicx}        % standard LaTeX graphics tool
\usepackage{multicol}        % used for the two-column index
\usepackage[bottom]{footmisc}% places footnotes at page bottom

\usepackage{cite}
\usepackage{subcaption}
\usepackage{newtxtext}       % 
\usepackage[varvw]{newtxmath}       % selects Times Roman as basic font

\newcommand{\old}[1]{{
	\color{red} \footnotesize [\textbf{\textsf{Old}}: \textsf{#1}]
	}}
\newcommand{\new}[1]{{
	\color{blue}  [\textbf{\textsf{New}}: \textsf{#1}]
	}}

\makeindex             % used for the subject index
\begin{document}

\title*{Holographic approach to anomalous transport in a massive $U(1)$ gauge theory}
\author{Nishal Rai$^{1,2}$ and Eugenio Meg\'{\i}as$^{3}$}%\orcidID{0000-1111-2222-3333} and\\ Eugenio Meg\'{\i}as\orcidID{1111-2222-3333-4444}}
\institute{Nishal Rai \at $^1$Center for Astrophysics Gravitation and Cosmology (CAGC), SRM University Sikkim, Upper Tadong, Sikkim, India \\ $^2$Department of Physics, SRM University Sikkim, Upper Tadong, Sikkim, India.\\ \email{nishalrai10@gmail.com}
  \and Eugenio Meg\'{\i}as \at $^{3}$ Departamento de F{\'\i}sica At\'omica, Molecular y Nuclear and Instituto Carlos I de F{\'\i}sica Te\'orica y Computacional, Universidad de Granada, Avenida de Fuente Nueva s/n, E-18071 Granada, Spain}
\maketitle

\abstract{In this study, we explore a massive \(U(1)\) gauge holographic model with pure gauge and mixed gauge-gravitational Chern-Simons terms. By considering the full backreaction of the gauge field on the metric tensor, we explore the vortical and energy transport sectors. Our findings for the chiral vortical conductivity, $\sigma_V$, and the chiral magnetic/vortical conductivity of energy current show that $\sigma^\varepsilon_B = \sigma^\varepsilon_V$. Notably, we highlight a contribution to \(\sigma_V\) induced by the mixed term in the massive theory, which is absent in the massless case.
}

\section{Introduction}
AdS/CFT correspondence \cite{Maldacena:1997re,Witten:1998qj} states that, in the low energy limit, the large-$N_c$, ${\mathcal N}$ = 4 super Yang-Mills field theory in four-dimensional space is equivalent to the type IIB string theory in $AdS_5 \times S^5$  space.  It has become a very powerful theoretical tool for studying strongly coupled systems. It has a wide application in different fields of physics, such as condensed matter physics, hydrodynamics and QCD \cite{Grieninger:2023myf,Baggioli:2023tlc,Morales-Tejera:2024uzg, Landsteiner:2022wap, Rai:2023nxe, Mukhopadhyay:2020tky, Rai:2019vrr,Mukhopadhyay:2017ltk, Mukhopadhyay:2017hgc,Ammon:2021pyz,Grieninger:2021zik,Ahn:2024ozz,Chu:2024dti,Landsteiner:2015lsa,Landsteiner:2012dm,Megias:2013joa,Newman:2005as,Newman:2005hd,Megias:2016ery,Landsteiner:2013aba}  to name a few.  Our study will be focused on the application of the fluid/gravity correspondence in the context of hydrodynamics, in particular the study of the anomalous transport in this kind of systems.

Quantum chiral anomalies, arising in relativistic field theories of chiral fermions beyond perturbation theory, play a crucial role in relativistic hydrodynamics~\cite{book1,book2,book3,Landsteiner:2012kd}. Since the 1980s, anomaly-induced transport mechanisms have been widely studied~\cite{Vilenkin}. Key effects include the chiral magnetic effect, where a charge current is induced parallel to an external magnetic field~\cite{Fukushima:2008xe}, and the chiral vortical effect, where a vortex in a charged fluid generates a current~\cite{Kharzeev:2007tn,Banerjee:2008th,Son:2009tf,Landsteiner:2011cp}. These effects may be detectable in heavy ion collisions at RHIC and LHC~\cite{STAR:2009wot} and can lead to anomalous transport properties in materials like Weyl semimetals~\cite{Basar:2013iaa,Landsteiner:2013sja}. 
In this work, we explore and investigate the chiral vortical effects in a massive $U(1)$ gauge holographic model. To achieve this, we have accounted for the full backreaction of the gauge field on the metric tensor and included a mixed gauge-gravitational Chern-Simons term as well. To evaluate the transport coefficients we will be using the Kubo formalism where the anomalous conductivities are given by the following relations

	\begin{equation}
	\begin{split}
		\sigma_V&=\lim_{k_c\rightarrow 0}\dfrac{i}{2 k_c}\sum_{a,b} \epsilon_{abc}\langle J^aT^{0b}\rangle\vert_{\omega = 0}\,, \\
		\sigma_V^\varepsilon&=\lim_{k_c\rightarrow 0}\dfrac{i}{2 k_c}\sum_{a,b} \epsilon_{abc}\langle T^{0a}T^{0b}\rangle\vert_{\omega = 0} \,,
	\end{split}
	\label{eqku1}
\end{equation}
where $\sigma_V$ is the chiral vortical conductivity, and $\sigma^\varepsilon_V$ the chiral vortical conductivity of energy current, respectively. The chiral magnetic conductivities for charge, $\sigma_B$, and energy, $\sigma^\varepsilon_B$, current are given by
\begin{equation}
	\begin{split}
		\sigma_B &= \lim_{k_c\rightarrow 0}\dfrac{i}{2 k_c}\sum_{a,b} \epsilon_{abc}\langle J^aJ^{b}\rangle\vert_{\omega = 0} \,,\\
		\sigma_B^\varepsilon &= \lim_{k_c\rightarrow 0}\dfrac{i}{2 k_c}\sum_{a,b} \epsilon_{abc}\langle T^{0a}J^{b}\rangle\vert_{\omega = 0}\,.
	\end{split}
	\label{eqku2}
\end{equation}
	These correlators can be evaluated using the AdS/CFT dictionary \cite{Landsteiner:2011iq,Son:2002sd,Herzog:2002pc}.
\section{Holographic model}
\label{sec:2}
% Always give a unique label
% and use \ref{<label>} for cross-references
% and \cite{<label>} for bibliographic references
% use \sectionmark{}
% to alter or adjust the section heading in the running head

	The holographic action for a massive $U(1)$ gauge boson  with pure gauge and a mixed gauge-gravitational Chern-Simon term is given by ~\cite{Landsteiner:2011iq,Jimenez-Alba:2014iia,Rai:2023nxe}
\begin{eqnarray}
	S &=& \dfrac{1}{16 \pi G}\int d^5x\sqrt{-g}
	\Big[
	R+2\Lambda-\dfrac{1}{4} F_{MN} F^{MN} \nonumber \\
	&&-\dfrac{m^2}{2} (A_M-\partial_M\theta) (A^M-\partial^M\theta) \nonumber \\
	&&+ \epsilon^{MNPQR} (A_M-\partial_N\theta)\left(\dfrac{\kappa}{3}F_{NP}F_{QR}+\lambda R^A\,{}_{BNP}R^B\,{}_{AQR} \right)\Big] \nonumber \\
	&&+ S_{\rm GH}+S_{\rm CSK}  \,,  
	\label{act}
\end{eqnarray}
where $ S_{\rm GH}$ and $S_{\rm CSK} $
are the Gibbons-Hawking boundary term, and a boundary term induced by the mixed gauge-gravitational anomaly, respectively. $\theta$ is a field which ensures gauge invariance (up to gauge anomalies), and thus the mass term enters in a consistent way. As it mentioned in~\cite{Klebanov:2002gr,Gursoy:2014ela,Casero:2007ae}, the St\"uckelberg term arises as the holographic realization of dynamical anomalies. For convenience we will define a new field $B_M \equiv A_M-\partial_M\theta$, such that $\theta$ does not appear explicitly in the equations of motion. One may check that the action still remains invariant under gauge transformations. From now onwards we will be working with the field $B_\mu$ instead of $A_\mu$. 

\subsection{Background solution}
\label{subsec:2}
The ansatz for the background metric and  fields is chosen as follows in Fefferman-Graham coordinates ~\cite{Megias:2017czr,Megias:2019djo},
\begin{equation}
	\begin{aligned}
	  ds^2 &=-\dfrac{\ell^2}{\rho} g_{\tau\tau}(\rho) d\tau^2+\dfrac{\ell^2 }{\rho} g_{xx}(\rho)d\vec{x}^2+\dfrac{\ell^2 }{4\rho^2} d\rho^2 \,,\\
	  B_\mu dx^\mu &= B_t (\rho) dt \,, 
	\label{bckmet}
	\end{aligned}
\end{equation}
where the boundary lies at $\rho=0$ and the horizon at $\rho_h$, with  $g_{\tau\tau}(\rho_h)=0$. Using this ansatz the equations of motion for the background are given by 
	\begin{eqnarray}
	g_{xx}''(\rho )-\dfrac{g_{xx}'(\rho )}{\rho }    + \frac{1}{6\ell^2 \rho} \dfrac{ g_{xx}(\rho )}{g_{\tau \tau }(\rho )}\left(\dfrac{m^2 \ell^2
	}{4 } B_t(\rho )^2+\rho^2 B_t'(\rho )^2\right)
	&=&0 \,, \nonumber\\
	g_{\tau \tau }'(\rho ) \left(1-\rho \dfrac{g_{xx}'(\rho)}{g_{xx}(\rho )} \right) + g_{\tau \tau }(\rho ) \dfrac{g_{xx}^\prime(\rho )}{g_{xx}(\rho )} \left(3 - \rho\dfrac{  g_{xx}^\prime(\rho
		)}{g_{xx}(\rho )}\right)\label{eq:eom_B} \\
- \frac{1}{3} \dfrac{\rho ^2}{\ell^2} B_t^\prime(\rho)^2 + \dfrac{1}{12} m^2 B_t(\rho )^2&=&0\,, \nonumber\\
	B_t^{\prime\prime}(\rho ) + \frac{1}{2} \left( 3 \dfrac{g_{xx}^\prime(\rho )}{g_{xx}(\rho )}-\dfrac{g_{\tau\tau}^\prime(\rho )}{g_{\tau \tau }(\rho )}\right) B_t^\prime(\rho ) -\dfrac{\ell^2 m^2}{4 \rho^2}  B_t(\rho ) &= &0 \,. \nonumber
\end{eqnarray}
%The asymptotic behaviour of the background gauge field is 
%	\begin{equation}
%	B_M(\rho)=a_0 \rho ^{-\frac{\Delta}{2} }+    a_1 \rho ^{\frac{\Delta }{2}+1} + \cdots \,, \label{eq:B_asymptotic}
%\end{equation} 
%%% THIS NEEDS TO BE REWRITTEN
%where $m^2 \ell^2=\Delta(\Delta+2)$, with $\Delta$ the anomalous dimension of the dual current~\cite{Jimenez-Alba:2014iia}. 

%The first(second) term in Eq.~(\ref{eq:B_asymptotic}) corresponds to a non-normalizable(normalizable) mode.  The scaling dimension of the normalizable mode is  ($3+\Delta$), and this puts an upper bound on the value $\Delta=1$. For $\Delta>1$ the dual operators become irrelevant (in the IR), and so we will be working in the range of values of $\Delta$ below this bound. 

We will solve numerically the above equations of motion (\ref{eq:eom_B}) subjected to the following boundary conditions
\begin{equation}
	B_t(\rho_h)=0 \,, \hskip2cm \lim_{\rho \to 0} \left( \rho^{\Delta/2} B_{t}(\rho) \right) = \mu_5 \,,
\end{equation}
where $\mu_5$ corresponds to the source of the gauge field and $\Delta$ being the anomalous dimension of the dual current \cite{Jimenez-Alba:2014iia} related to the mass of the gauge field through the following relation $m^2 \ell^2=\Delta(\Delta+2)$. In addition to this, we demand the solution to the metric tensor to be regular at the horizon.

\begin{figure}[t]
	          \centering
        \begin{subfigure}{0.45\textwidth}
		\includegraphics[width=1\linewidth]{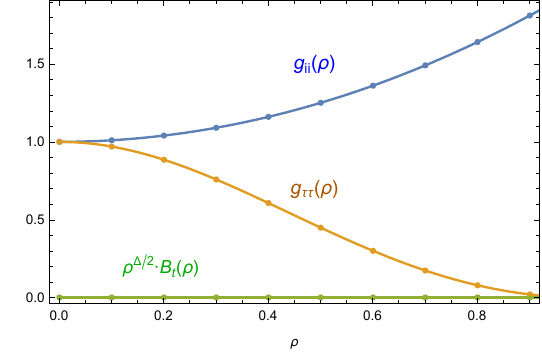}
	\end{subfigure}\hfill 
	\begin{subfigure}{0.45\textwidth}
		\centering
		\includegraphics[width=1\linewidth]{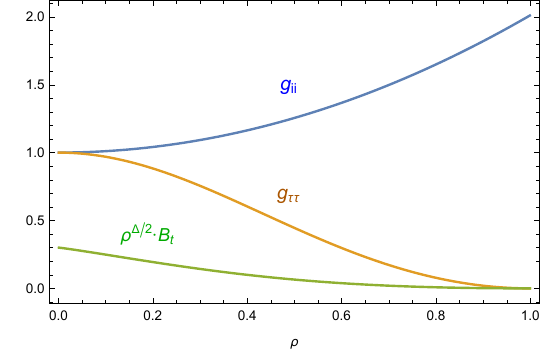}
	\end{subfigure}
		\caption{(color) Dependence of the background metric and gauge field with~$\rho$. We display the results for $g_{xx}(\rho)$ (blue), $g_{\tau\tau}(\rho)$ (orange),  and  $\rho^{\Delta/2} B_{t}(\rho)$ (green). Left panel: We have considered $\mu_5 = 0$, with solid line being the analytic solution and the dots the numerical one. Right panel: We have chosen $\Delta=0.1$ and $\mu_5=0.3$.}
	\label{fback}
\end{figure}

It is possible to find an analytic solution of the background for the case with $\mu_5 = 0$. This is given by
\begin{equation}
	g_{\tau\tau}(\rho) = \frac{1}{\rho_h^2} \frac{(\rho_h^2 - \rho^2)^2}{\rho_h^2 + \rho^2} \,, \qquad g_{xx}(\rho) = 1 + \frac{\rho^2}{\rho_h^2}  \,, \qquad B_t(\rho) = 0 \,.
\end{equation}
We display in Fig.~\ref{fback} (left) the numerical solution of the equations of motion for $\mu_5=0$ (dots), and compare it with the corresponding analytical solution (solid lines). We display also in the right panel the numerical solution for $\mu_5 = 0.3$.

\section{Anomalous conductivity}
To compute these correlators, we will follow the general procedure of the perturbation of the fields on top of the background, and later use the AdS/CFT dictionary to compute the correlators. 

\subsection{Fluctuations} 
Within this approach, we perturb the fields on top of the numerical solution for the background displayed in Fig.~\ref{fback}. We will study the linear response of the fluctuations by splitting the metric and gauge field into background and linear perturbation components, i.e.
\begin{equation}
	g_{MN}=g_{MN}^{(0)}+\epsilon h_{MN}, \hskip2cm     B_{M}=B_{M}^{(0)}+\epsilon b_{M} \,,
	\label{eqpert}
\end{equation}
where the terms with ${}^{(0)}$ correspond to the background metric/field, and the terms with $\epsilon$ the fluctuations of this metric/field. Then, we will follow the general procedure of Fourier mode decomposition \cite{Amado:2011zx}
\begin{eqnarray}
	h_{MN}(\rho, x^\mu) &=& \int \dfrac{d^dk}{(2\pi)^d}h_{MN}(\rho)e^{-i\omega t+i \vec{k}.\vec{x}} \,,\\
	b_{M}(\rho, x^\mu) &=& \int \dfrac{d^dk}{(2\pi)^d}b_{M}(\rho)e^{-i\omega t+i \vec{k}.\vec{x}}\,.
\end{eqnarray}
Without loss of generality, we consider perturbations with frequency \(\omega\) and momentum \(k\) in the \(z\)-direction. 
Since we aim to compute correlators at zero frequency, we set the frequency-dependent parts to zero in the equations and solve the system up to first order in \(k\). In this limit, the fields \(h^i_z\) decouple from the system and become constant. Thus, we can write the system of differential equations for the shear sector as follows
\begin{eqnarray}
	&&b_i^{\prime\prime}(\rho )+    \frac{1}{2} \left(\frac{g_{xx}^\prime(\rho )}{g_{xx}(\rho
		)}+\frac{g_{\tau \tau }^\prime(\rho )}{g_{\tau \tau }(\rho )}\right) b_i^\prime(\rho )  -\frac{\Delta  (\Delta
		+2)}{4 \rho ^2}  b_i(\rho )  \label{fleq1}\\
	&&+\left(\frac{4 i \kappa  k \epsilon_{ij} b_j(\rho )}{\sqrt{g_{xx}(\rho ) g_{\tau \tau}(\rho )}}+\frac{g_{xx}(\rho )
		h^i{}_t^\prime(\rho )}{g_{\tau \tau }(\rho )}\right)B_t'(\rho )+  i \lambda k\epsilon_{ij}h^j{}_t'(\rho ) \Omega(\rho) =0,\nonumber\\
	&&h^i{}_t^{\prime\prime}(\rho ) - \left(\dfrac{g_{\tau \tau }'(\rho )}{2
		g_{\tau \tau }(\rho )} - \dfrac{5 g_{xx}'(\rho )}{2 g_{xx}(\rho
		)} + \dfrac{1}{\rho }\right) h^i{}_t'(\rho )+\dfrac{\rho B_t'(\rho )}{g_{xx}(\rho )}  b_i^\prime(\rho ) \nonumber\\
	&&+\dfrac{\Delta  (\Delta +2)  B_t(\rho )}{4 \rho  g_{xx}(\rho )} b_i(\rho ) + i \lambda k\epsilon_{ij} \Phi_j(\rho) =0 \,,\label{fleq2}
\end{eqnarray}
where $i, j = x, y$. The explicit expressions of the functions $\Omega(\rho)$ and $\Phi_j(\rho)$ are given in Appendix A of \cite{Rai:2023nxe}.

The asymptotic behaviour of the fields up to the first subleading term are given by 
	\begin{eqnarray}
	b_i(\rho) &=& b_i^{(0)} \rho ^{-\frac{\Delta}{2} }+    b_i^{(1)} \rho ^{\frac{\Delta }{2}+1} + \cdots \,,\\
	h_{t}^i(\rho) &=& h_{t}^i{}^{(0)} +    h_{t}^i{}^{(1)} \rho^2 + \cdots  \,,
\end{eqnarray}
where the leading-order terms \(b_i^{(0)}\) and \(h_{t}^i{}^{(0)}\) play the role of sources.

\subsubsection{Conductivities}
From the holographic description of the correlation functions, one can evaluate the one-point functions as
\begin{equation}
\langle J_a\rangle=\dfrac{\delta S_{\text{ren}}}{\delta b_a^{(0)}}=-\dfrac{2}{16\pi G}(\Delta+1)b_a^{(1)}\,,  \quad
\langle T_{0a}\rangle=\dfrac{\delta S_{\text{ren}}}{\delta h_{t}^a{}^{(0)}}=\dfrac{1}{16\pi G}\left(2h_{t}^a{}^{(0)}+h_{t}^a{}^{(1)}\right) \,,
\end{equation} 
with $ (a = x,y)$. $S_{\textrm{ren}} = S + S_{\textrm{ct}}$ is the renormalized action, with $S$ the action given in Eq.~(\ref{act}) and $S_{\textrm{ct}}$ the counterterm. The two-point functions can be obtained by taking the variation of one-point function with respect to the corresponding source term and the conductivities are evaluated using the Kubo formulae given in (\ref{eqku1}) and (\ref{eqku2}), i.e.
	\begin{align}
	\sigma_V &= \lim_{k\to 0} \frac{1}{k} \textrm{Im} \langle J_x T_{0y}\rangle \,, \qquad \sigma_V^\varepsilon  = \lim_{k\to 0} \frac{1}{k} \textrm{Im} \langle T_{0x} T_{0y}\rangle \,, \\
	\sigma_B &= \lim_{k\to 0} \frac{1}{k} \textrm{Im} \langle J_x J_y \rangle \,, \qquad\hspace{0.1cm} \sigma_B^\varepsilon = \lim_{k\to 0}\frac{1}{k} \textrm{Im} \langle T_{0x} J_y \rangle \,.
\end{align}
A comparison of the variation of the action under axial gauge transformations with  the consistent form of the anomaly for chiral fermions~\cite{book1}  allows us to set the value of the anomaly coefficients to   
\begin{equation}
	\kappa = -\frac{G}{2\pi} \,, \qquad \lambda = -\frac{G}{48\pi} \,.
	\label{ano-coe}
\end{equation}
	\begin{figure}[h]
	\centering
	\begin{subfigure}{0.45\textwidth}
		\centering
		\includegraphics[width=1\linewidth]{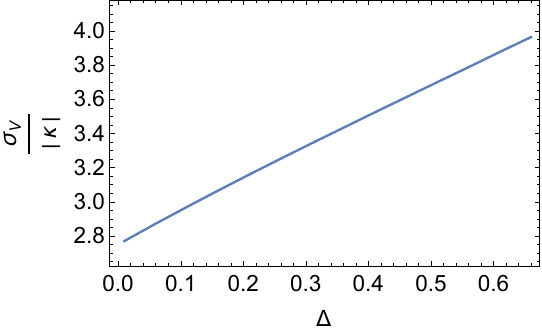}
	\end{subfigure}\hfill
	\begin{subfigure}{0.45\textwidth}
		\includegraphics[width=1\linewidth]{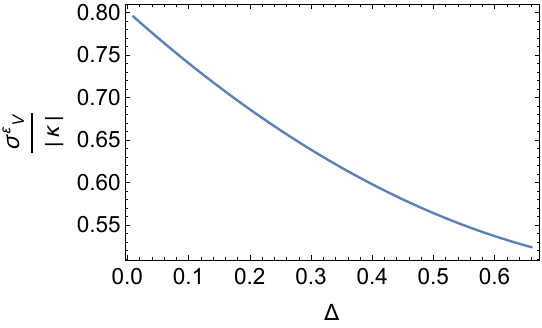}
	\end{subfigure}
	\centering
	\begin{subfigure}{0.45\textwidth}
		\centering
		\includegraphics[width=1\linewidth]{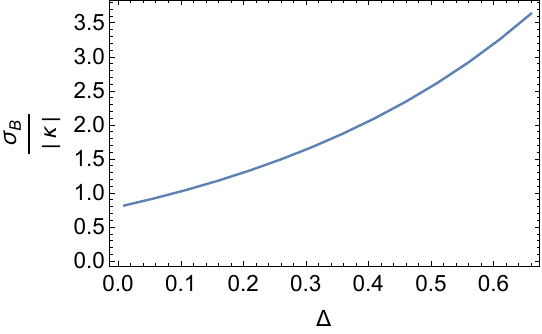}
	\end{subfigure}\hfill
	\begin{subfigure}{0.45\textwidth}
		\includegraphics[width=1\linewidth]{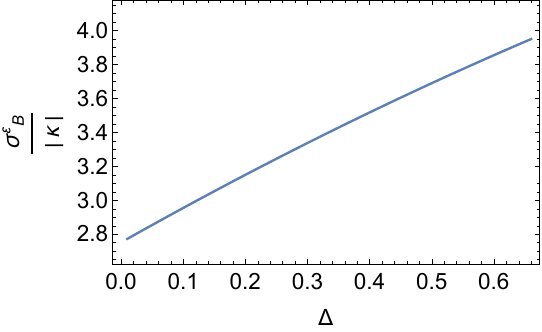}
	\end{subfigure}
	\caption{Upper panels: Plot of $\sigma_V$ (left) and $\sigma^\varepsilon_V $ (right) vs $\Delta$. Lower panels: Plot of $\sigma_B$ (left) and  $\sigma^\varepsilon_B $ (right) vs $\Delta$. We have considered $\mu_5= 0.15$ in all the panels. We have divided the conductivities by $|\kappa|$ to make them dimensionless. }
	\label{figsg1}
\end{figure}
        In this regard we will be setting $G = 2\pi$, so that $\kappa=-1$ and $\lambda=-1/24$. In Fig.~\ref{figsg1} we have plotted the anomalous conductivities vs $\Delta$ (a parameter related to the mass of the gauge field). From this figure, we observe that the chiral vortical conductivity and the chiral magnetic conductivity for energy current are the same even at finite mass, i.e., \(\sigma_V = \sigma^\varepsilon_B\), and both increase with \(\Delta\). Additionally, we see that the chiral vortical conductivity of energy current, \(\sigma^\varepsilon_V\), decreases with \(\Delta\), but the rate decreases rapidly. Conversely, in the case of the chiral magnetic conductivity, \(\sigma_B\), it increases with \(\Delta\) as shown in Fig.~\ref{figsg1}. 
In  the limit of vanishing mass, our results lead to $\frac{\sigma_B}{\mu_5 |\kappa|}\simeq 16/3$, which exactly coincides with the results in  \cite{Landsteiner:2011iq,Jimenez-Alba:2014iia} where $\alpha$ has been set to $\mu_5$ in both references~\footnote{$\alpha$ corresponds to the asymptotic value of the gauge field $A_t$ for $\rho \to 0$. In our case, we assume $\alpha = \mu_5$ for $\Delta = 0$.}.
\section{Discussion}

In this work we have studied the mass dependence of the anomalous conductivities in a holographic model for a massive \(U(1)\) gauge field, incorporating both pure gauge and mixed gauge-gravitational anomaly terms. In this model, the vortical sector was also accessible, and the conductivities corresponding to this sector were also studied. We observed that the chiral vortical conductivity, \(\sigma_V\), and the chiral magnetic conductivity for energy current, \(\sigma^\varepsilon_B\), are equal and increase with \(\Delta\). An intriguing finding is the presence of contributions to \(\sigma_B\)   induced by the mixed gauge-gravitational anomaly term in the massive theory, which were absent in the massless case. Furthermore, the conductivities \(\sigma_B\), \(\sigma_V\), and \(\sigma^\varepsilon_B\) all increase with \(\Delta\), while the chiral vortical conductivity of energy current, \(\sigma^\varepsilon_V\), decreases with \(\Delta\). We have explicitly verified that all our numerical results for the conductivities at finite mass converge to the known results at zero mass as \(\Delta \to 0\).

\section*{Acknowledgments}
N.R. thanks the International Centre for Theoretical Sciences (ICTS) for the support through the program - Field Theory and Turbulence (ICTS/FTT2023/12). The works of N.R. and E.M. are supported by the project PID2020-114767GB-I00 and by the Ram\'on y Cajal Program under Grant RYC-2016-20678 funded by MCIN/AEI/10.13039/501100011033 and by ``FSE Investing in your future'', as well as by the FEDER/Junta de Andaluc\'{\i}a-Consejer\'{\i}a de Econom\'{\i}a y Conocimiento 2014–2020 Operational Program under Grant A-FQM-178-UGR18. The work of E.M. is also supported by Junta de Andaluc\'{\i}a under Grant FQM-225, and  by the ``Pr\'orrogas de Contratos Ram\'on y Cajal'' Program of the University of Granada.

%%%%%%%%%%%%%%%%%%%%%%%% referenc.tex %%%%%%%%%%%%%%%%%%%%%%%%%%%%%%
% sample references
% %
% Use this file as a template for your own input.
%
%%%%%%%%%%%%%%%%%%%%%%%% Springer-Verlag %%%%%%%%%%%%%%%%%%%%%%%%%%
%
% BibTeX users please use
% \bibliographystyle{}
% \bibliography{}
%
%\biblstarthook{

\end{document}